\documentstyle[12pt]{article}
\begin{document}
\baselineskip 5mm

\newcommand{\ov}{\overline}
\newcommand{\r}{$^{[??]}$}
\newcommand{\pa}{\partial}
\newcommand{\bops}{\mathop{\textstyle\otimes}}

\makeatletter
\def\eqnarray{%
 \stepcounter{equation}%
 \let\@currentlabel=\theequation
 \global\@eqnswtrue
 \global\@eqcnt\z@
 \tabskip\@centering
 \let\\=\@eqncr
 $$\halign to \displaywidth\bgroup\@eqnsel\hskip\@centering
 $\displaystyle\tabskip\z@{##}$&\global\@eqcnt\@ne
 \hfil$\displaystyle{{}##{}}$\hfil
 &\global\@eqcnt\tw@$\displaystyle\tabskip\z@{##}$\hfil
 \tabskip\@centering&\llap{##}\tabskip\z@\cr}
\makeatother

\title{ The Fuzzy K\"ahler Coset Space with \\ 
  the Darboux Coordinates }

\author{Shogo Aoyama\thanks{e-mail: spsaoya@ipc.shizuoka.ac.jp}\ \ \ \ and 
\ \ Takahiro Masuda\thanks{e-mail: stmasud@ipc.shizuoka.ac.jp} \\
      \\
       Department of Physics \\
              Shizuoka University \\
                Ohya 836, Shizuoka  \\
                 Japan}
                 
\maketitle

\begin{abstract}
 The Fedosov deformation quantization of the symplectic manifold is determined by a 
$1$-form differential $r$. We identify a class of $r$ for which the $\star$ product becomes 
the Moyal product by taking  appropriate   Darboux coordinates, but invariant 
 by canonically transforming the  coordinates. 
 This respect of the $\star$ product is explained by studying the fuzzy algebrae of the 
K\"ahler coset space. 
\end{abstract}

\vspace{6cm}



\newpage

The discovery of non-commutative quantum field theories in the M theory and the string 
theory\cite{0} revived an old idea of non-commutativity of the spacetime in physics, and 
gave rise to the intensive current activity among the field theorists.( See for instance 
\cite{1} for the recent review and references therein.)
 Their consideration  was mainly focused on non-commutativity in the flat spacetime. No 
doubt the final goal is to  study   non-commutative quantum field theory in a curved 
spacetime. One of the approaches  towards to this direction\cite{2,4} is the deformation 
quantization of the symplectic manifold by the Fedosov formalism\cite{4}.

\vspace{1cm}

 The Fedosov deformation quantization is determined by the $1$-form differential $r$, which  
is defined as a solution to eq. (\ref{m4}).  It  depends on many things, i.e.,   local 
coordinates of the symplectic manifold, a symplectic connection $\Gamma$, a local frame 
$\theta$ and an initial condition $\mu$, as can be seen from eq. (\ref{m5}).  But any 
solution $r$ defines a non-commutative $\star$ product which satisfies the associativity. 
It is suspected that for a class of the  $1$-form differential $r$ the $\star$ product 
would become  simple and have invariance by some coordinate transformations. In this letter 
we discuss the issue in the Darboux coordinates. We identify the class of $r$ (see eq. 
(\ref{sol}) for which  the $\star$ product   reduces to the Moyal product by retaking an 
appropriate set of the Darboux coordinates.  It then follows that the Moyal product thus 
obtained is invariant by any canonical transformation of the Darboux coordinates. 

\vspace{1cm}

As an application we study the fuzzy algebrae for the K\"ahler coset space $G/H$ in the 
Darboux coordinates. In ref. \cite{6} it was shown that the Killing potentials satisfy 
in the holomorphic coordinates the fuzzy algebrae 
\begin{eqnarray}
 [ M^A(z,\ov z), M^B(z,\ov z) ]_\star
 &=& -i(c_1 \hbar + c_3\hbar^3 + c_5\hbar^5 + \cdots)\sum_{A=1}^{dim\ G}f^{ABC} M^C(z,\ov 
z),  \label{eq 0} \\
\sum_{A=1}^{dim\ G} M^A(z,\ov z)\star M^A(z,\ov z) &=& R  + c_2\hbar^2 + c_4\hbar^4 + 
\cdots\ ,  \label{eq 00}
\end{eqnarray}
when the coset space is irreducible. 
The coefficients $c_1,c_2,c_3, \cdots $ are numerical constants. In this letter we find 
these coefficients as $c_i = 0$ for $i\ge 3$,  working in a particular set of the Darboux 
coordinates. The invariance of  the Moyal product, stated above, implies that this simple 
form of the fuzzy algebrae remains to be the same in whichever set of  the Darboux 
coordinates we work.

\vspace{1cm}

We start with reviewing on the Fedosov construction in the generalized form\cite{4,5}. 
Consider a real $2N$-dimensional symplectic manifold $\cal M$ with local coordinates $(x^1,x^2,\cdots,x^{2N})$. The symplectic $2$-form is given by
\begin{eqnarray}
\omega ={1\over 2} \omega_{ij}dx^i \wedge dx^j.  \nonumber
\end{eqnarray} 
We introduce a local frame of $T^*\cal M$ given by an isomorphism
\begin{eqnarray}
dx^i \longrightarrow  \theta^i(x) \ . \nonumber
\end{eqnarray} 
The local $1$-forms $\theta^i$ are not necessarily closed ($d\theta^i \ne 0$). In this 
local frame $\omega$ is transported to 
\begin{eqnarray}
\Omega_0 ={1\over 2} \omega_{ij}d\theta^i \wedge d\theta^j.  \label{m1}
\end{eqnarray}
We  think of deforming  a $q$-form differential $a(x)$  as formal power series:
\begin{eqnarray}
a(x) \longrightarrow a(x;y) = 
 \sum^{\infty}_{p=0}{1\over p!q!}a(x)_{i_1i_2\cdots i_pj_1j_2\cdots j_q}
  y^{i_1}y^{i_2}\cdots y^{i_p}\theta^{j_1}\theta^{j_2}\cdots \theta^{j_q}\ ,
  \label{m2}
\end{eqnarray}
where $a_{j_1j_2\cdots j_q }(x) = a(x)$, and $(y^1,y^2,\cdots,y^{2N})$ are  deformation 
coordinates. For such deformed differentials the $\circ$ product is defined by 
\begin{eqnarray}
 a(x;y) \circ b(x;y) &=& \sum_n {1\over n!}(-{i\hbar \over 2})^n \omega^{i_1 
j_1}\omega^{i_2 j_2}
\cdots \omega^{i_n j_n}\partial_{i_1}^y\partial_{i_2}^y\cdots \partial_{i_n}^ya
 \partial_{j_1}^y\partial_{j_2}^y\cdots \partial_{j_n}^yb. \label{circ}
\end{eqnarray}
The $\star$ product is induced from the $\circ$ product    as
\begin{eqnarray}
a(x)\star b(x) = a(x;y)\circ b(x;y)|_{y=0} \ .  \label{m2'}
\end{eqnarray}
We also define the covariant derivative $\partial$ and the form-changing operators $\delta$ 
and $\delta^{-1}$:
\begin{eqnarray}
\partial a  =  da + [\Gamma, a ]_{\circ} \ ,    \quad \quad\quad
\delta a  =  \theta^i {\partial \over \partial y^i}a = [\omega_{ij}y^i\theta^j, a ]_{\circ}  
\ ,  \nonumber  \\   
\delta^{-1} a_{pq}  =  {1\over p+q}y^i {\partial \over \partial \theta^i} a_{pq}  
\ ,  \hspace{3cm} \nonumber   
\end{eqnarray}
where $\Gamma$ is the symplectic connection of the manifold ${\cal M}$ and $a_{pq}$ is the 
part of degree $p$ in $y$ and order $q$ in $\theta$ of (\ref{m2}).

\vspace{1cm}

The deformation (\ref{m2}) is determined so as to obey the constraint 
\begin{eqnarray}
Da \equiv \partial a - \delta a + {i\over \hbar }[r, a]_{\circ} = 0 \ ,  \label{m3}
\end{eqnarray}
in which $r$ is a $1$-form satisfying 
\begin{eqnarray}
\delta r = \partial (\omega_{ij}y^i  \theta^j ) + {\cal R}  + \partial r + {i\over 
\hbar}r\circ r \ . \label{m4}
\end{eqnarray}
Eq. (\ref{m4}) is a sufficient condition to guarantee $D^2 a = 0$.  It has a unique 
solution obeying $ \delta^{-1} r = \mu  $.
It can be shown 
 by iterating  the equation in the form
\begin{eqnarray}
r = \delta \mu + \delta^{-1} [\ \partial (\omega_{ij}y^i  \theta^j ) + {\cal R}  + \partial 
r + {i\over \hbar}r\circ r \ ] \ . \label{m5}
\end{eqnarray}
Once a  solution for $r$ given, the deformation  (\ref{m2}) is  explicitly found by solving  
the constraint (\ref{m3}). The solution is unique in the case where  $a(x;y)$ is a  
$0$-form differential.

\vspace{1cm}

According to the Darboux theorem there exist local coordinates in the neighborhood of any 
point $x \in {\cal M}$, called the Darboux coordinates, such that 
\begin{eqnarray}
\omega &=& dp_1\wedge dq^1 + \cdots\cdots + dp_N\wedge dq^N.  \label{m6} 
\end{eqnarray}
In the first place we study deformation quantization by simply choosing the local frame as 
given by 
\begin{eqnarray}
(\theta^1,\theta^2,\cdots,\theta^{2N}) = (dp_1,\cdots,dp_N,dq^1,\cdots,dq^N) \ , \label{m7} 
\end{eqnarray}
with which  $\Omega_0 = \omega $.
In the Darboux coordinates one can take the symplectic connection $\Gamma$ to vanish
so that ${\cal  R} = 0$. Then eq. (\ref{m5}) has the trivial  solution  $r=0$ by choosing   
 $\mu = 0$. Solving the constraint (\ref{m5}) with $r=0$  we find the unique deformation
\begin{eqnarray}
a(p,q) \longrightarrow  a(p,q;\xi,\zeta) = a(p+\xi,q+\zeta)\ , \label{m7'}
\end{eqnarray}
for a $0$-form differential. 
Here  $N$-tuples of $\xi$ and $\zeta$ are deformation coordinates in the local frame 
(\ref{m7}).
Then the $\star$ product (\ref{m2'}) reduces to the ordinary Moyal product.
The Darboux coordinates are not unique. We may have 
\begin{eqnarray}
\omega &=& dp'_1\wedge dq'^1 + \cdots\cdots + dp'_N\wedge dq'^N \ ,  \nonumber
\end{eqnarray}
by a canonical transformation 
\begin{eqnarray}
(p,q) \longrightarrow (p'(p,q),q'(p,q)) \ . \label{ccc}
\end{eqnarray}
It is evident that the above arguments hold in any of these coordinates.

\vspace{1cm}

Next we study the deformation quantization in the local frame where 
\begin{eqnarray}
\Omega_0 &=& \theta_1\wedge \theta^1 + \cdots\cdots + \theta_N\wedge \theta^N \ , \nonumber 
\end{eqnarray}
with an isomorphism
\begin{eqnarray}
\theta_\alpha = f_\alpha^\beta dp_\beta+g_{\alpha\beta}dq^{\beta} \ , \quad\quad    
\theta^\alpha = h^{\alpha\beta} dp_\beta+ j_{\beta}^{\alpha}dq^\beta \ . \quad\quad\quad   
\label{m8}  
\end{eqnarray}
Here  $f_\alpha^{\beta}, g_{\alpha\beta},h^{\alpha\beta}$ 
and $j_\alpha^\beta$  are
 local functions of the Darboux coordinates $p$ and $q$. We assume the isomorphism to be 
symplectic, i.e., 
\begin{eqnarray}
f_{\alpha}^\beta j_{\beta}^{\gamma}&-& g_{\alpha\beta
}h^{\beta\gamma}=\delta^{\gamma}_{\alpha}, \nonumber \\
f^{\alpha}_{\beta}h^{\beta\gamma}&-&f^{\gamma}_{\beta}
h^{\beta\alpha}=j^{\beta}_{\alpha}g_{\beta\gamma}
-j^{\beta}_{\gamma}g_{\beta\alpha}=0.  \label{m8'}
\end{eqnarray}
Then  $\Omega_0$ is equal to $\omega$ again. But the local frame (\ref{m8}) cannot be 
related to (\ref{m7}) by a canonical transformation since $d\theta^\alpha \ne 0$ and 
$d\theta_\alpha \ne 0$ generically. One may wonder if taking this local frame 
would yield us  a different deformation quantization. To examine this we first of all  
solve (\ref{m5}) which now reads  
\begin{eqnarray}
r = \delta \mu + \delta^{-1} [\ d (\xi_\alpha\theta^\alpha- 
\zeta^\alpha\theta_\alpha ) + d r + {i\over \hbar}r\circ r \ ] \ , \label{m9}
\end{eqnarray}
by $ \Gamma = 0$ and ${\cal R}= 0$. 
This time we have 
$$
d (\xi_\alpha\theta^\alpha- \zeta^\alpha\theta_\alpha ) \ne 0 \ .
$$  
For any choice of $\mu$ (\ref{m9}) can be solved by iteration. However we expect that by a 
clever choice  the solution is given by
$$
r = \delta \mu + \delta^{-1} [ d(\xi_\alpha\theta^\alpha- \zeta^\alpha\theta_\alpha 
)]  \ , 
$$
satisfying 
\begin{eqnarray}
d r + {i\over \hbar}r\circ r  = 0 \   \label{sol}.
\end{eqnarray}
This indeed happens when we choose $\mu$ as
$$
\mu=
A^{\beta\gamma}_{\delta}\,\xi_{\beta}\xi_{\gamma}\zeta^{\delta}
+B_{\beta\gamma}^{\delta}\,\xi_\delta\zeta^{\beta}\zeta^{\gamma}
+C^{\beta\gamma\delta}\,\xi_{\beta}\xi_{\gamma}\xi_{\delta}
+D_{\beta\gamma\delta}\,\zeta^{\beta}\zeta^{\gamma}\zeta^{\delta} \ ,
$$
with 
\begin{eqnarray}
A^{\beta\gamma}_{\delta} &=& {1\over 3}(\pa_{\delta}h^{\beta\gamma} 
-\pa^{\beta}j^{\gamma}_{\delta} ) 
+{1\over 2}
(j^{\beta}_{\sigma}\tilde{\pa}_{\delta}h^{\sigma\gamma}
 -h^{\sigma\beta}\tilde{\pa}_{\delta}j^{\gamma\sigma})  \ ,\nonumber  \\
B_{\beta\gamma}^{\delta}&=& {1\over 3}( \pa_{\beta}f_{\gamma}^{\delta}
-\pa^{\delta}g_{\beta\gamma})
-{1\over 2}(g_{\beta\sigma}\tilde{\pa}^{\delta}j^{\sigma}_{\gamma}
-f^{\sigma}_{\beta}\tilde{\pa}^{\delta} g_{\sigma\gamma}) \ ,\nonumber \\
C^{\beta\gamma\delta}&=&
{1\over 6}(-j^{\gamma}_{\sigma}\tilde{\pa}^{\delta}h^{\sigma\delta}
+h^{\sigma\gamma}\tilde{\pa}^{\delta}j^{\delta}_{\sigma}) \ , \label{x} \\
D_{\beta\gamma\delta}&=&{1\over 6}
( g_{\gamma\sigma}\tilde{\pa}_{\delta}f^{\sigma}_{\delta}
-f^{\sigma}_{\gamma}\tilde{\pa}_{\delta}g_{\sigma\delta}) \ , \nonumber
\end{eqnarray}
by using the notation
$$
\tilde{\pa}_{\delta}=g_{\delta\rho}\,{\pa\over \pa p_{\rho}}-
f_{\delta}^{\rho}\,{\pa \over \pa q^{\rho}}, \ \ \  
\tilde{\pa}^{\delta}=j^{\delta}_{\rho}\,{\pa\over \pa p_{\rho}}
-h^{\delta\rho}\,{\pa\over \pa q^{\rho}}.
$$
The solution takes the form 
\begin{eqnarray}
r={1\over 2} \{(h^{\beta\gamma}dj_{\gamma}^{\delta}
-j_{\gamma}^{\beta}dh^{\delta\beta})
\xi_{\gamma}\xi_{\delta}
+(f_{\beta}^{\delta}dj_{\gamma}^{\beta}-g_{\gamma\beta}dh^{\delta\beta})
\xi_\delta\zeta^\gamma
 + (g_{\gamma\beta}df_{\delta}^{\beta}-
f_{\gamma}^{\beta}dg_{\delta\beta})\zeta^{\gamma}\zeta^{\delta} \} . \label{r}
\end{eqnarray}
 Using this solution for $r$ we solve  (\ref{m3}) to get the deformation in the form 
\begin{eqnarray}
 a(p,q) \longrightarrow a(p,q;\xi,\zeta) = a(P(p,\xi),Q(q,\zeta)) \ . \nonumber
\end{eqnarray}
After calculations we find the solution for $P$ and $Q$ in the simple forms
\begin{eqnarray}
P_\alpha (p,\xi,\zeta) = p_\alpha + j_{\alpha}^{\beta} \xi_\beta
   -g_{\beta\alpha}\zeta^{\beta}
   \ ,   \quad\quad
Q^\alpha (q,\xi,
\zeta) = q^\alpha -h^{\beta\alpha}\xi_{\beta}+f_\beta^\alpha \zeta^\beta 
  \ .   \label{m10}
\end{eqnarray}
It is interesting to compare this deformation with (\ref{m7'}), the one obtained in the 
local frame (\ref{m7}). We observe  that  the deformation coordinates are transformed by 
\begin{eqnarray}
\xi_\alpha  \longrightarrow  \ \ j_{\alpha}^{\beta} \xi_\beta
   -g_{\beta\alpha}\zeta^{\beta} \ , 
  \quad\quad
 \zeta^\alpha  \longrightarrow   -h^{\beta\alpha}\xi_{\beta}+
f_\beta^\alpha \zeta^\beta  \ ,   \label{simp} 
\end{eqnarray}
which is a symplectic transformation due to (\ref{m8'}). 
Therefore  the $\circ$ product (\ref{circ}) in the local frame (\ref{m8}) 
 reduces to the one defined in the original frame (\ref{m7}). So does the $\star$ 
product. When $d\theta = 0$, 
 these local frames are  related with each other 
by a canonical transformation of the Darboux coordinates such as (\ref{ccc}). 
In other words, when $d\theta = 0$, for the class of $r$ given by (\ref{r})  the $\star$ 
product becomes the Moyal product  and invariant by any canonical transformation of the 
Darboux coordinates.

\vspace{1cm}

We shall give a concrete example for the Darboux coordinates in the case where ${\cal M}$ 
is the K\"ahler manifold. The K\"ahler manifold has local complex coordinates $z^\alpha 
=(z^1,z^2,\cdots ,z^N)$ and their complex conjugates. The symplectic 2-form 
reduces to the K\"ahler $2$-form given by 
$$
\omega = ig_{\alpha\ov \beta}dz^\alpha\wedge d\ov z^\beta = i {\partial^2 K \over \partial 
z^\alpha \partial \ov z^\beta } dz^\alpha\wedge d\ov z^\beta \ .
$$
It can be put in the form 
\begin{eqnarray}
\omega = dp_\alpha \wedge dq^\alpha \ , \label{m11}
\end{eqnarray}
with
\begin{eqnarray}
p_\alpha = i{\partial K \over \partial \ov z^\alpha } \ , \quad\quad\quad
 q^\alpha = \ov z^\alpha \ .   \label{m12}
\end{eqnarray}
Hence they are the  Darboux coordinates. 

\vspace{1cm}

It is interesting to study the fuzzy K\"ahler coset space $G/H$ in these Darboux 
coordinates, and examine the fuzzy algebrae of the Killing potentials.
 To this end we have to remind of the method for constructing the K\"ahler coset space 
$G/H$\cite{9}. We consider the irreducible case. Then the group $G$ have generators $T^A 
=\{ X_\alpha,\ov X^\alpha, H^i,Y \}$   which satisfy the Lie-algebra
\begin{eqnarray}
[ X_\alpha, \ov X^\beta ] &=&  t(\Gamma^i)_\alpha^\beta H^i +
 s \delta_\alpha^\beta Y,  \quad\quad  [ X_\alpha, X_\beta ] \ =\  0,  \nonumber \\
 \quad [ X_\alpha, H^i ] &=& (\Gamma^i)_\alpha^\beta X_\beta,\quad\quad [X_\alpha, Y]\ =\  
X_\alpha, \quad c.c.,   \label{eq 36'}
\end{eqnarray}
with some constants $t$ and $s$ depending on the representation of $G$.
Here $X_\alpha$ and $\ov X^\alpha$ are coset generators. In the method of ref. \cite{9} the 
local coordinates of $G/H$ are denoted by $z_\alpha$ and $\ov z^\alpha$, where upper or 
lower indices stand for complex conjugation. Therefore raising or lowering tensor indices 
should be  done by writing the metrics $g_\alpha^{\ \beta}$ or $(g^{-1})_\alpha^{\ \beta}$ 
explicitly. Simple algebra gives 
\begin{eqnarray}
[X_\alpha, [X_\gamma, \ov X^\beta ] ] = \{t (\Gamma^i)_\alpha^\beta(\Gamma^i)_\gamma^\delta 
+ s \delta_\alpha^\beta \delta_\gamma^\delta)\} X_\delta 
  \equiv  M_{\alpha\gamma}^{\beta\delta} X_\delta.  \label{eq 37}
\end{eqnarray}
The quantity $M_{\alpha\gamma}^{\beta\delta} $ plays a key role in the method and has a 
remarkable property. It is 
summarized by the statement that
\begin{eqnarray}
M_{\alpha_0\alpha_1}^{\gamma_1\beta_1}M_{\alpha_2\beta_1}^{\gamma_2\beta_2}\cdots
M_{\alpha_{n-1}\beta_{n-2}}^{\gamma_{n-1}\beta_{n-1}}
M_{\alpha_{n}\beta_{n-1}}^{\gamma_{n}\beta_{n}}          \label{eq 39}
\end{eqnarray}
is completely symmetric in the indices $(\gamma_1,\cdots,\gamma_n,\beta_n)$, whenever  it 
is completely symmetrized in the indices $(\alpha_0,\alpha_1,\cdots,\alpha_n )$, and vice 
versa. 
The Killing vectors  $R_{A\alpha}\ (\ov R^{A \alpha})$ of the coset apace $G/H$ are 
non-linear realizations of the Lie-algebra (\ref{eq 36'}) on $z_\alpha\ (\ov z^\alpha) $:
\begin{eqnarray}
R^A_{\ \alpha} \equiv -i[T^A, z_\alpha], \quad c.c.,   \label{eq 38} 
\end{eqnarray}
which  are given by
 \begin{eqnarray}
R^\gamma_{\ \alpha} &=& i\delta^\gamma_\alpha, \quad\quad 
R_{\gamma\alpha} \ =\ {i\over 2}M_{\alpha\gamma}^{\beta\delta} z_\beta z_\delta,
 \nonumber  \\
R^i_{\ \alpha} &=&  i(\Gamma^i)_\alpha^\beta z_\beta, \quad\quad
R_\alpha \ =\ i z_\alpha.   \nonumber
\end{eqnarray}
The K\"ahler potential takes the form
\begin{eqnarray}
K(z,\ov z) = \ov z {1\over Q}\log (1+Q)z,   \label{1}       
\end{eqnarray}
where the semi-positive definite matrix $Q_\alpha^\beta$ is defined by
\begin{eqnarray}
Q_\alpha^\beta = -{1\over 2} M_{\alpha\gamma}^{\beta\delta}\ov z^\gamma z_\delta.
 \nonumber
\end{eqnarray}
By expanding the logarithm in (\ref{1}) in powers of $Q$ it can be shown that 
\begin{eqnarray}
K(z,\ov z) &=& \ov z (1 - {1\over 2}Q + {1\over 3}Q^2 - \cdots\cdots ) z \nonumber \\
   &=& \ov z z - {1\over 2!}[X_\alpha,z_\beta ] \ov z^\alpha \ov z^\beta +
    {1\over 3!} [X_\alpha,[X_\beta,z_\gamma ]]\ov z^\alpha \ov z^\beta \ov z^\gamma \cdots 
\ .
     \label{3}   
\end{eqnarray}
The last line follows upon using the symmetry property of (\ref{eq 39}).

\vspace{1cm}

We take a particular normalization of the Lie-algebra (\ref{eq 36'}) such that
$t=s=-1$. 
We then find the explicit form of the Killing potentials $M^A$:
\begin{eqnarray}
K_\alpha &=& ({1\over 1+Q}z)_\alpha,   \quad\quad\quad\quad\quad 
\ov K^\alpha = (\ov z {1\over 1+Q})^\alpha ,  \label{no} \\
M^i &=& \ov K\Gamma^i z \ =\  \ov z \Gamma^i K,  \quad\quad\quad  
M = \ov K z -1 \ =\  \ov z K -1 \ .  \nonumber     
\end{eqnarray}
In the third equation use was made of  the formula
$
\ov z\Gamma^i Q^n z = \ov z Q^n \Gamma^i z . 
$
They indeed transform according to (\ref{eq 36'})  under the Lie-variation. It suffices to 
show the transformations
\begin{eqnarray}
{\cal L}_{R_\gamma} K_\alpha &=&  0, \quad\quad\quad 
{\cal L}_{R_\gamma} \ov K^\alpha = i[(\Gamma^i)^\alpha_\gamma M^i + \delta_\gamma^\alpha M 
]\ .  \label{9}
\end{eqnarray}
Other transformations trivially follow from these. The Lie-variations in  (\ref{9}) can be 
written with the commutator defined by (\ref{eq 38}):
\begin{eqnarray}
{\cal L}_{R_\gamma} K_\alpha =  -i[X_\gamma,K_\alpha ],  \quad\quad 
{\cal L}_{R_\gamma} \ov K^\alpha =  -i[X_\gamma,\ov K^\alpha ]\ .  \label{9'}
\end{eqnarray}
Note the formulae for $K_\alpha$ and $\ov K^\alpha$:
\begin{eqnarray}
K_\alpha &=& {\partial \over \partial \ov z^\alpha}K 
   = z_\alpha - [X_\alpha, z_\beta]\ov z^\beta + 
   {1\over 2!} [X_\alpha,[X_\beta,z_\gamma ]] \ov z^\beta \ov z^\gamma - \cdots\cdots  \ , 
   \label{9'''} \\
\ov K^\alpha &=& \ov z^\alpha + {1\over 2}M^{\alpha\beta}_{\gamma\delta}\ov z^\gamma \ov 
z^\delta K_\beta \ .    \label{10} 
\end{eqnarray}
The last formula follows  by calculating as 
\begin{eqnarray}
{1\over 2}M^{\alpha\beta}_{\gamma\delta}\ov z^\gamma \ov z^\delta K_\beta &=&
 {1\over 2}M^{\alpha\beta}_{\gamma\delta}\ov z^\gamma \ov z^\delta 
 ({1\over 1+Q}z)_\beta 
 = {1\over 2}(\ov z {1\over 1+Q})^\gamma M^{\alpha\beta}_{\gamma\delta}\ov z^\gamma
 \ov z^\delta \ov z_\beta \ , \nonumber
\end{eqnarray}
with recourse to the symmetry property of the multiple product (\ref{eq 39}). 
Calculating the commutators in (\ref{9'}) by 
 (\ref{9'''} ) and (\ref{10})  we obtain (\ref{9}). Finally we may check that
\begin{eqnarray}
M^AM^A &=& 2\ov K^\alpha K_\alpha  + M^iM^i \ +\  MM\   \nonumber \\
   &=& 2\ov z ({1\over 1+Q})^2 z \ +\  \ov K\Gamma^i z \cdot \ov z\Gamma^i K \ +\ 
    \ov K z \cdot \ov z K \ -\  2 \ov K z \ +\ 1   
    = 1.  \nonumber
\end{eqnarray}

\vspace{1cm}

By using (\ref{9'''}) we find that the Darboux coordinates are given by
$$
p_\alpha = i K_\alpha \ , \quad\quad\quad 
q^\alpha = \ov z^\alpha \ , 
$$
according to (\ref{m12}) and (\ref{9'''}). In terms of the Darboux coordinates
 the Killing potentials in (\ref{no}) take the forms
\begin{eqnarray}
\ov K^\alpha &=& q^\alpha - {i\over 2}M_{\gamma\delta}^{\alpha\beta}q^\gamma q^\delta 
p_\beta \ ,  \quad\quad \quad
K_\alpha = -i p_\alpha \ ,    \nonumber  \\
M^i &=& -i q\Gamma^i p \ , \quad\quad\quad\quad\quad 
M = -i q p -1 \ ,  \nonumber  
\end{eqnarray}
where use is made of   (\ref{10}). In the Darboux coordinates we may choose the ordinary 
Moyal product as the $\star$ product, as has been discussed. 
Little calculation shows that these Killing potentials $M^A (p,q)$ satisfy the fuzzy 
algebrae
\begin{eqnarray}
 [ M^A(p,q), M^B(p,q) ]_\star 
&=& -i \hbar f^{ABC} M^C(p,q),  \label{11} \\
   \nonumber  \\
 M^A(p,q)\star M^A(p,q) &=& 1 -{\hbar^2 \over 2}(tr \Gamma^i\Gamma^i + N)  \ , \label{12}
\end{eqnarray}
which are much simpler than (\ref{eq 0}) and (\ref{eq 00}). 
 This calculation may be done in any other set of the Darboux coordinates, say $p'$ and 
$q'$. The transformation of the coordinates induces the symplectic isomorphism of the local 
frame as given by (\ref{m8}) and (\ref{m8'}). According  to the arguments which followed in 
that paragraph, 
 the quantum deformations of $M^A(p',q')$ and $M^A(p,q)$ in the respective local frames  
$(dp',dq')$ and $(dp,dq)$  can be related  by the symplectic transformation (\ref{simp}).
 Therefore the simple form of the fuzzy algebrae (\ref{11}) and (\ref{12}) are invariant by 
the transformation (\ref{ccc}).

\vspace{1cm}

Finally we apply the whole arguments in this letter  for $CP^N (= U(N+1)/U(N)\otimes U(1) )
$. The generators of $U(N+1)$ are decomposed as $T^A = \{X_\alpha,\ov X^\alpha, 
H_\alpha^\beta, Y \}$. 
They satisfy the Lie-algebra (\ref{eq 36'}) with $(H_\alpha^\beta)_\gamma^\delta = 
-\delta_\alpha^\delta \delta_\gamma^\beta $ and $t=s=-1$. 
The quantity $M_{\alpha\gamma}^{\beta\delta}$, defined by (\ref{eq 37}), takes the form
$$
M_{\alpha\gamma}^{\beta\delta} = -\delta_\alpha^\beta \delta_\gamma^\delta - 
\delta_\alpha^\delta \delta_\gamma^\beta \ .
$$
Then we find the K\"ahler potential 
$K = \log (1+ \ov z z )$ 
from (\ref{3}) and  the Killing potentials\cite{10}
\begin{eqnarray}
K_{\alpha}&=&{z_{\alpha}\over 1+\ov{z}z},\ \ \ \ \ \quad \quad 
{\ov K}^{\alpha}={\ov{z}^{\alpha}\over 1+\ov{z}z},\nonumber \\
M_\alpha^\beta &=&{ z_\alpha \ov z^\beta \over 
1+\ov{z}z}\ , \ \ \ \quad\quad  M=-{1\over 1+\ov{z}z} \ ,     \label{kill}
\end{eqnarray}
from (\ref{no}).
The Darboux coordinates are given by
\begin{eqnarray}
p_{\alpha}= i {z_{\alpha}\over 1+\ov{z}z},\ \ q^{\alpha}=\ov{z}^{\alpha}, \nonumber
\end{eqnarray}
with which the Killing potentials (\ref{kill}) are expressed as
\begin{eqnarray}
K_{\alpha} &=&-i p_\alpha ,\ \ \ \ \ \quad \quad 
{\ov K}^{\alpha}= q^\alpha (1 + iqp )  ,\nonumber \\
M_\alpha^\beta &=& -i p_\alpha q^\beta  , \ \ \ \quad\quad  M=-iqp -1 \ . \nonumber  
\end{eqnarray}
We may be interested in the real coordinates 
\begin{eqnarray}
p'_{\alpha}={\ov{z}^{\alpha}z_{\alpha}\over 1+\ov{z}z}, \ \ \ \ \ \quad 
q'^{\alpha}={1\over 2i}\log{z_{\alpha}\over \ov{z}^{\alpha}} \ , \quad\quad\quad
 (\hbox{no sum over} \ \alpha) \ , \nonumber
\end{eqnarray}
which are regarded as radial and angle coordinates for a fixed $\alpha$. 
They are also the Darboux coordinates because 
$$
 \sum_{\alpha=1}^N dp'_\alpha\wedge dq'^\alpha =  \sum_{\alpha=1}^N dp_\alpha\wedge 
dq^\alpha \ .
$$
Both Darboux coordinates are related by a canonical  transformation such that 
\begin{eqnarray}
p'_{\alpha}= -i q^\alpha p_\alpha  \ , \ \ \ \ \ 
q'^{\alpha}={1\over 2i}[ \log {p_{\alpha}\over q^{\alpha}} - \log (1 + iqp)
 - {\pi \over 2}i ] \ ,   \label{cano}
\end{eqnarray}
or  
\begin{eqnarray}
p_{\alpha}=\sqrt{(P'-1)p'_{\alpha}}\,e^{iq'^{\alpha}}\ , \ \ \ \ \ \quad\quad\quad 
q^{\alpha}=\sqrt{{ p'_{\alpha}\over 1-P'} }\,e^{-iq'^{\alpha}}    \ ,  \nonumber
\end{eqnarray}
with no sum over $\alpha$ and  $P' = \sum_\alpha p'_\alpha$.
 The Killing potentials may be rewritten as 
\begin{eqnarray}
K_{\alpha}&=&\sqrt{(1- P')p'_{\alpha}}\,e^{iq'^{\alpha}},\ \ \ \ \ \
\ov{K}^{\alpha}=\sqrt{(1-P')p'_{\alpha}}\, 
e^{-iq'^{\alpha}}  \ , \nonumber\\
M_\alpha^\beta &=& \sqrt{p'_{\alpha}p'_{\beta}}
\,e^{i(q'^{\alpha}-q'^{\beta})},\ \ \ \ \ \  M= -(1-P') \ , \quad\quad\quad
    \label{new} 
\end{eqnarray}
with no sum over $\alpha$ and $\beta$. 
We study the fuzzy algebrae of the Killing potentials (\ref{new}) with the Moyal product in 
the coordinates
 $p'$ and$q'$. Obviously the canonical transformation (\ref{cano}) induces a symplectic  
isomorphism between $(dp,dq)$ and $(dp',dq')$ as given by (\ref{m8}). 
 Therefore 
 the  Killing potentials satisfy the fuzzy algebrae  (\ref{11}) and (\ref{12}) in the 
coordinates $p'$ and $q'$ as well.
 We have also checked this by a direct calculation with (\ref{new}). 
 
 \vspace{1cm}
 
For  $CP^1$ we may furthermore transform the local frame $(dp',dq')$ to 
$(\theta^1,\theta^2)$ as
\begin{eqnarray}
\theta^1 = {1\over r}dp' \ , \quad\quad\quad\quad 
\theta^2 = rdq' \ ,  \label{cp}  \label{kisi}
\end{eqnarray}
where $ r = \sqrt{\ov z z}$. The transformation is symplectic so that the fuzzy algebrae 
(\ref{11}) and (\ref{12}) still remain to be the same even in the local frame 
(\ref{kisi}). 
In ref. \cite{5'} the deformation quantization of $CP^1$ was discussed in this local frame 
and the same fuzzy algebrae were obtained. However note that (\ref{cp}) does not induce a 
canonical transformation of the 
Darboux coordinates $(p',q')$ at all, because $d\theta^2 \ne 0$.
        
\vspace{1cm}

In this letter we have identified the class of the $1$-form differential $r$ for which   
the $\star$ product naturally reduces to the ordinary Moyal product by taking appropriate 
Darboux coordinates $(p,q)$. It was  shown that the Moyal product thus obtained is 
invariant in whichever local frame $\theta$  we work, as long as the isomorphism $(dp,dq) 
\rightarrow \theta$ is symplectic. When $d\theta = 0$, the 
isomorphism  is nothing but a canonical transformation of  the Darboux coordinates $(p,q)$. 
Owing to this invariance of the Moyal product we were able to show
 that for  the irreducible K\"ahler coset space $G/H$  the  Killing potentials satisfy the 
fuzzy algebrae  (\ref{11}) and (\ref{12}) in any set of Darboux coordinates. It is 
desirable to generalize the arguments for the reducible case\cite{11}.

\vspace{2cm}
\noindent
{\Large\bf Acknowledgements}

One of the authors (T.M.) thanks T. Asakawa, I. Kishimoto and S. Watamura  for the 
discussions. 
His work  was supported in part by JSPS Research Fellowships for Young Scientists.

\hspace{3cm}

\end{document}